\documentclass[primary,nonacm]{acmart}
\renewcommand\footnotetextcopyrightpermission[1]{} 

\usepackage{multirow}
\usepackage{multicol}
\usepackage{hyperref}
\usepackage{tabularx}
\usepackage{graphicx}
\usepackage{xcolor}
\usepackage{mathtools}
\usepackage{amsmath}
\usepackage{caption}

\title{Multi-Queue SSD I/O Modeling \& Its Implications for Data Structure Design}
\author{Erin Ransom}
\email{erins.ransom@gmail.com}
\affiliation{\institution{Harvard University} \country{United States}}
\author{Andrew Lim}
\email{andrewlim@alumni.harvard.edu}
\affiliation{\institution{Harvard University} \country{United States}}
\author{Michael Mitzenmacher}
\email{michaelm@eecs.harvard.edu}
\affiliation{\institution{Harvard University} \country{United States}}

\begin{abstract}
Understanding the performance profiles of storage devices and how best to utilize them has always been non-trivial due to factors such as seek times, caching, scheduling, concurrent access, flash wear-out, and garbage collection. 
However, analytical frameworks that provide simplified abstractions of storage performance can still be accurate enough to evaluate external memory algorithms and data structures at the design stage.  
For example, the Disk Access Machine (DAM) model assumes that a storage device transfers data in fixed-size blocks of size $B$ and that all transfers have unit latency.  
This abstraction is already sufficient to explain some of the benefits of data structures such as B-trees and Log-Structured Merge trees (LSM trees); however, storage technology advances have significantly reduced current models' accuracy and utility. 

This paper introduces the Multi-Queue Solid State Drive (MQSSD) model, a new storage abstraction.
This model builds upon previous models and aims to more accurately represent the performance characteristics of modern storage hardware.    
We identify key performance-critical aspects of modern multi-queue solid-state drives on which we base our model and demonstrate these characteristics on actual hardware.  
We then show how our model can be applied to LSM-tree-based storage engines to optimize them for modern storage hardware.  
We highlight that leveraging concurrent access is crucial for fully utilizing the high throughput of multi-queue SSDs, enabling designs that may appear counterintuitive under traditional paradigms 
We then validate these insights through experiments using Facebook's LSM-tree-based key-value store, RocksDB.   
We conclude that the MQSSD model offers a more accurate abstraction of modern hardware than previous models, allowing for greater insight and optimization. 

\end{abstract}

\begin{document}

\maketitle

\section{Introduction}
\label{sec:intro}

Understanding the performance of storage devices and the implications for data structure design is challenging due to factors such as seek times, caching, scheduling, parallelism, flash wear-out, and garbage collection.
To address this complexity, researchers and designers have long used analytical frameworks such as the Disk Access Machine (DAM), Parallel Disk Access Machine (PDAM), and Affine models, which provide a simplified view of I/O performance but capture enough performance-critical features to evaluate the relative performance of different structures and design choices \cite{Aggarwal88, Affine}.  
For example, the DAM model considers only the number of pages transferred between main memory to storage, but this was sufficient to explain the benefits of B-trees and their rise to prominence in the 1970s \cite{btree}, as well as the more recent prevalence of write-optimized structures such as $B^e$-trees and Log-Structured Merge-trees (LSM-trees) \cite{betrfs, lsm}.
The PDAM and Affine models refine the DAM model by accounting for parallelism and seek time, respectively.  
These minor refinements improve predictions for different types of storage hardware, such as Hard Disk Drives (HDDs) and Solid State Drives (SSDs), and have provided greater insight into design choices such as node size in B-trees vs. $B^e$-trees \cite{refinements}.   
  
However, advances in storage technology have outpaced our abstractions for I/O, leading to substantial gaps in the insight they can provide.  
For example, the move from mechanical addressing in HDDs to electrical addressing and out-of-place updates in SSDs has significantly reduced overall latency, but it has also introduced new overheads related to writes and garbage collection to manage flash wear-out, resulting in asymmetric read and write performance \cite{mqsim}.
The address translations required for out-of-place updates also significantly impact performance.  The order in which pages are accessed directly impacts the efficiency of the devices' Flash Translation Layer (FTL) \cite{mqsim}.
Modern multi-queue SSDs are also designed to use Peripheral Component Interconnect Express (PCIe) and support much higher transfer speeds and overall throughput than previous Serial AT Attachment (SATA) HDDs and SSDs \cite{sata,  pcie_spec, pcie}.
However, the highly parallel architectures of these devices require high levels of concurrent access to fully utilize their potential, as reflected by their use of the Non-Volatile Memory Express (NVMe) protocol, which provides multiple I/O queues directly accessible by applications \cite{nvme}.
The resulting performance profile for a multi-queue SSD is then sensitive to several factors, including concurrency, order of access, and write intensity.
I/O models that overlook these factors will naturally be less accurate and provide less insight when evaluating storage algorithms.

To illustrate this, consider a basic file I/O benchmark designed to demonstrate the effects of concurrency and order of access on latency and throughput. 
We use $k$ threads to read $k$ \texttt{GB} of data concurrently from a 50 \texttt{GB} file, each thread reading 1 \texttt{GB} of data using $r$ equally sized sequential reads starting at staggered random offsets, resulting in $r$ random accesses per thread.
We then vary $r$ from $0.1\%$ to $100\%$ of pages read and $k$ from 1 to 128, resulting in $rk$ random page reads for a given trial.
We use staggered random offsets to ensure that no two threads read the same data simultaneously.  
Between the staggering and the file size, the vast majority of randomly accessed pages will result in cache misses for the device.

The DAM model would suggest that the total latency of this experiment is purely dependent on the total amount of I/O with no correlation to $r$ and that throughput will be constant with respect to $k$. 
Similarly, the PDAM model predicts no correlation with $r$ but does account for an increase in throughput proportional to the device's hardware-level parallelism, i.e., the number of physical channels to the device's flash backend.  
By contrast, the Affine model can reasonably predict the effects of random access for a single thread but does not capture how this changes when adding more threads.  
None of these models would predict any difference if we performed writes instead of reads.  
In practice, we observe that single-threaded throughput varies by as much as $100$x with respect to $r$ across multiple devices and upwards of $26x$ for writes. 
Moreover, we see this relationship change as we increase $k$ and throughput that scales well beyond the level of hardware parallelism.  
When we add to this substantially different performance profiles for writes vs reads, the limits of these I/O models become rather glaring. 

Despite this, the need for storage algorithms optimized for such hardware is only increasing. 
Multi-queue SSDs have become increasingly prevalent, gaining widespread adoption in the cloud and in enterprise deployments - not only are they cheap and fast, but recent advancements in the NVMe-over-Fabric protocol have made disaggregated flash storage even more tractable \cite{nvme_alibaba, nvme_of}. 
However, existing database systems and storage engines are not designed to exploit the performance of multi-queue SSDs fully \cite{nvme_perf_gap}. 
Closing this gap will require new storage algorithms and paradigms, which our current I/O abstractions are not equipped to assess.  

In this paper, we present the Multi-Queue SSD (MQSSD) model, which can capture the impacts of concurrency and random access on modern hardware with respect to both reads and writes.  
We build off the knowledge of existing I/O models, as described in Section \ref{sec:background:models} and base our new model on the performance critical characteristics of multi-queue SSDs.
We outline these characteristics and their performance implications in Section \ref{sec:background:mqssds} and demonstrate the proficiencies and shortcomings of previous models when applied to multi-queue SSDs. 
The comparison of these models described earlier can be seen in Figure \ref{fig:model_comp} along with the MQSSD model. 
In Section \ref{sec:definition}, we define the MQSSD model and validate its accuracy using the same benchmarks. 
In Section \ref{sec:applications}, we demonstrate how the MQSSD model can be used to evaluate and optimize a real-world storage engine designed for flash storage, Facebook's RocksDB.
When compared against a configuration used by Facebook for high-throughput query workloads, prioritizing concurrency can improve query throughput by a factor of $2$x without sacrificing write throughput.
In Section \ref{sec:related_work}, we compare our work against other related projects. 
In Section \ref{sec:conclusion}, we conclude that the MQSSD model provides greater accuracy and insight than previous I/O models when designing and optimizing storage systems for modern hardware; however, there is still room for improvement.  

\section{Background }
\label{sec:background}

We now describe in detail the existing I/O models we build upon and aspects of multi-queue SSDs we want our model to capture.  

\subsection{Current I/O Models}
\label{sec:background:models}
Introduced by Aggarwal and Vitter in 1988 \cite{Aggarwal88}, the Disk Access Machine (DAM) model, also known as the external memory or I/O model, has been a cornerstone of I/O analysis for decades.
The core assumptions of this model are that main memory and storage are primarily differentiated by their size, access granularity, and access latency, which remains true of storage hardware to this day.  
However, even minor refinements can significantly enhance the model’s accuracy, as we will see when we consider the PDAM and Affine models.

\subsubsection{The DAM Model}
The DAM model captures the fundamental relation between any two levels of the memory hierarchy; however, we focus on the context of main memory and storage, for which the model was initially intended.
We have the following parameters:
\begin{itemize}
    \item[$N$:] The number of records/size of the working set.  Each record is assumed to have a unit size.  
    \item[$M$:] The number of records that can fit in main memory.   
    \item[$B$:] The number of sequentially stored records that can be transferred to or from storage in a single page, i.e., our granularity of access.  
\end{itemize}
In this model, our cost unit is page transfers, i.e., reading a page into memory or writing it back to storage, the idea being that the latency of a single page transfer is on a different order of magnitude than in-memory computation.  
For instance, modern Dynamic Random Access Memory (DRAM) has access latencies measured in nanoseconds, compared to milliseconds for contemporary HDDs \cite{ddr5, hdd}. 
Due to their different orders of magnitude in latency, in-memory computation is considered free in this context.  
It also assumes that $ 1 \leq B \leq M < N $.

\subsubsection{The PDAM Model}
The PDAM model is a variation of the DAM model also introduced by Aggarwal and Vitter in 1988 \cite{Aggarwal88} that was designed to represent HDDs with some level of concurrent access with multiple read/write heads, e.g., RAID0. 
To account for this, the PDAM adds the following parameter to the DAM:
\begin{itemize}
    \item[$P$:] The number of pages that can be transferred concurrently. 
\end{itemize}
Since this model allows for multiple transfers at a time at no additional cost, the unit of cost is no longer individual page transfers.
Instead, we consider the number of I/O cycles during which up to $P$ pages can be concurrently transferred.   
If we fix $P=1$, this is then equivalent to the DAM model.  
The PDAM model also assumes $1 \leq P \leq \lfloor M/B\rfloor $.
More recent work by Bender et al. has shown that this model can reasonably predict SSD performance when $P$ is taken to be the number of physical channels to flash on the device in the context of concurrent-read-exclusive-write I/O \cite{refinements}.

\subsubsection{The Affine Model}
Introduced by Ruemmler and Wilkes in 1994 \cite{Affine}, the Affine model refines the DAM by taking into consideration the order in which pages are accessed, breaking down the cost of an I/O operation into two components based on how mechanical addressing handles sequential vs. random I/O:
\begin{itemize}
    \item[$s$:] Setup cost, i.e., the time required to reposition the HDDs read/write head for random access.  
    \item[$\alpha$:] Bandwidth cost, i.e., the time to transfer an individual page once the read/write head is in place.  This is the only cost for sequentially accessed pages.   
\end{itemize}
In this model, reading or writing $N$ consecutive records to or from disk costs $s + \frac{\alpha N}{B}$, $s$ for the initial random access and $ \alpha $ for each of the $N/B$ pages accessed.
The specific values for $s$ and $\alpha$ are device-dependent but can be derived through benchmarking and linear regression.  
This model captures the large discrepancy in latency between random and sequential transfers in HDDs.
Bender et al.\cite{refinements} show that $s$ is on the order of $\alpha \times 10^2$ for commodity HDDs. 
SSDs use electrical addressing vs. the mechanical addressing this model is based on; however, for reasons we will discuss in Section \ref{sec:background:mqssds}, this model still accurately reflects the relationship between random and sequential I/O in SSDs.

\subsection{Example: B-Trees} 
\label{sec:backgroud:ex}
To illustrate the use and differences of these models, we examine the B-tree, a widely used data structure for external storage.   
A B-tree is a balanced search tree on $N$ key-value pairs where all leaf nodes have the same depth, internal nodes consist of pivots and pointers, and the leaves store the key-value pairs. 
Each node is stored sequentially, but following a pointer from one node to another requires random access.  
We also assume that the leaves of the tree are connected as a linked list. 
To optimize this structure, we seek to minimize the cost of lookups, inserts, and updates in the respective model.
We also consider the \textit{data structure write amplification} of updates, which we define as the amortized amount of data written to storage per operation divided by the amount of actual data modified per update.  
Regardless of the model, the worst-case write amplification in a B-tree scales linearly with node size, as modifying a node requires rewriting it entirely.

\noindent \textbf{DAM Model:} 
Under the DAM model, where all page transfers are treated equally, the primary goal is to maximize the utility of each transfer.
When performing any operation in a B-tree other than updating the key-value pairs, the primary cost is traversing the tree, as updates to the internal nodes are relatively infrequent and cheap. 
The cost of reading or updating the key-value pairs will be the same regardless of the tree's shape. 
Using nodes of size $\Theta(B)$  maximizes the utility of each transfer as we retrieve or update the maximum number of records for a single transfer while minimizing the total number of transfers required to traverse the tree.  
Such a tree has depth $O(\log_B\frac{N}{B})$, and any operation on $X$ sequential records will cost $O(\log_B(\frac{N}{B}) + \frac{X}{B})$ transfers to traverse the tree and then perform said operation on the $\frac{X}{B}$ pages containing the records. 
Barring fundamental changes to the data structure, this also minimizes write amplification as $B$ is the smallest increment of access to storage. 
See Brodal and Fagerberg for the full proof of asymptotic optimality in the DAM model \cite{lower_bounds}.  

\noindent \textbf{PDAM Model:} 
For the PDAM model, where we can perform up to $P$ concurrent transfers in a cycle, we consider how to minimize the number of cycles required for the same operations.  
Again, we aim to maximize the utility of each cycle. 
As before, if we have $X$ consecutive records, the cost of actually operating on the records is still independent of the tree shape, but we can now perform our operations in batches of $P$ pages for a cost of $O(\frac{X}{PB})$ cycles. 
Moreover, if we use nodes of size $\Theta(PB)$, we can read or write an entire node in a single cycle, and we reduce the depth of the tree to $O(\log_{PB}\frac{N}{PB})$ for a total cost of $O(\log_{PB}(\frac{N}{PB}) + \frac{X}{PB})$.  
This also increases our write amplification by a factor of $P$; however, since we can write $P$ pages per cycle, this does not change the latency of a write in the PDAM model.  
The proof of asymptotic optimality is analogous to the proof for the DAM model, as the cost of a cycle vs. a page transfer is equivalent in the respective models.

\noindent \textbf{Affine Model:} 
Moving to the Affine model, we are back to considering individual pages, but with the distinction that we incur an additional setup cost, $s$, for random access instead of just the bandwidth cost, $\alpha$, for sequential access.  
Suppose we keep nodes of size $\Theta(B)$. 
In that case, our costs are essentially equivalent to those in the DAM model as reading or updating any given node will incur a cost of $\Theta(s + \alpha)$; however, if we use nodes of size $\Theta(YB)$ for an integer variable $Y$, then accessing a given node has a cost of $\Theta(s + Y\alpha)$.  
Since $s << \alpha $, the additional cost of accessing a node of size $\Theta(YB)$ is relatively low but reduces the depth of the tree to $O(\log_{YB} \frac{N}{YB})$.  
Therefore, using larger nodes allows us to reduce the number of random accesses for a given operation at the cost of additional sequential accesses. 

Unlike in the previous models, this has implications for the cost of traversing the tree and performing operations over our $X$ records, as we may need to traverse multiple leaf nodes, incurring additional random accesses.  
With nodes of size $\Theta(YB)$, an operation over $X$ consecutive records has a cost of $O((\log_{YB}(\frac{N}{YB}) + \frac{X}{YB})(s + Y\alpha))$ as we only pay the setup cost once for each node.  
As shown by Bender et al., the cost of traversing the tree is optimal when we set our node size to $\Theta(s/\alpha \ln (s/\alpha))$; however, if we are operating over multiple sequential records, such as a range query, the total cost will be asymptotically optimal with nodes of size $\Theta(s/\alpha)$\cite{refinements}. 
The general idea here is that large nodes amortize the setup cost incurred by random access; however, flattening the tree too much decreases its utility.  
There is then a tradeoff in performance between operations touching only a few records and operations covering many consecutive records. 
There is a similar tradeoff concerning write amplification, as write amplification in a B-tree will always be the worst for minor updates.  
The optimal configuration then ultimately depends not only on the values of $s$ and $\alpha$ but also on the type of workload.
This difference has been borne out in practice. 
Online Transaction Processing (OLTP) databases based on B-trees use small leaf nodes as they tend to have many small lookups and updates.  In contrast, Online Analytical Processing (OLAP) databases, which tend to favor large-range queries, typically use much larger leaf nodes \cite{azure}.

It is worth noting that there are other performance implications for B-tree node size outside the scope of these models. 
For example, if the database is caching its nodes in memory, then using larger nodes will reduce the total number that can fit in the cache, reducing its benefit.  
B-trees also have relatively poor write amplification regardless of node size, so small-update-intensive applications are more likely to use write-optimized structures like LSM trees, which we will discuss later.

\section{Multi-Queue Solid State Drives}
\label{sec:background:mqssds}
SSDs have become cheaper, larger, and longer lasting, so they have increasingly replaced HDDs in applications.  
Multi-queue SSDs using the NVMe protocol, in particular, have delivered unprecedented performance regarding latency and throughput and are particularly beneficial for I/O intensive applications \cite{nvme_perf}. 
Despite sharing some aspects with HDDs, such as page access granularity, SSDs store, read, and update data using fundamentally different mechanics. 
Therefore, the base framework of the DAM model is still relevant. 
However, as with PDAM and Affine models, we can improve the insight from our analysis by tailoring our model to the performance characteristics of the underlying hardware.  
Spinning platters with magnetic read/write heads have been replaced by hierarchical flash memory with electrical addressing. 
Instead of seek times, these devices must contend with out-of-place updates, wear leveling, and garbage collection \cite{flash_memory, steady_state, mqsim}.
This section describes the aspects of multi-queue SSDs that are performance-critical for I/O-intensive applications.  
For additional details on the internals of modern, multi-queue SSDs, we recommend the work of Tavakkol et al. \cite{mqsim}.

\subsection{Architecture}
We can divide the internals of a NAND-flash-based Multi-queue SSD into the \textit{back end} memory devices and the \textit{front end} control and management units.  
The front end includes the host-interface logic, which typically uses the NVMe protocol \cite{nvme}, and the Flash Translation Layer (FTL), which manages transactions over multiple bus channels, each connected to one or more flash chips \cite{mqsim}. 
These chips are organized in a highly hierarchical manner to maximize I/O concurrency. 
Each chip has a multiplexed interface over multiple dies that can independently execute commands over their respective planes \cite{flash_memory}.

Like with previous storage, the flash memory for these devices is also organized into pages, this being the minimum access granularity (typically 4 KB). 
However, a given page must be erased between each write due to how the flash cells are programmed. 
This means that each time a piece of data is updated, it must be moved to a new physical page on the device in a process referred to as \textit{out-of-place updates}.
Once the updated data is written to a new physical page, the old page is marked invalid and ready to be erased. 
Moreover, individual flash cells have limited lifespans due to wear caused by usage.
While reads cause minimal wear, writing and (especially) erasing take a significant toll due to the amount of charge required \cite{flash_memory}.   
Since data must be moved regardless, the FTL can send incoming writes to the least used pages, ensuring all pages wear at roughly the same rate in a process known as \textit{wear leveling} \cite{steady_state}.
Another wear-reducing measure is to erase pages in contiguous groups, referred to as blocks, as this reduces the total charge needed and the amount of wear on the flash cells \cite{flash_memory}.

While this combination of techniques has been quite effective at extending the longevity of SSDs, writing and erasing at two different granularities introduces complications.  
Ideally, the device can wait until a given block has been completely invalidated before it is erased; however, the user application ultimately determines when a given page is invalidated.  
If the number of free pages drops below a certain threshold, garbage collection will be triggered, and the device will start moving valid pages to free up more space.  
This results in \textit{hardware write amplification}, which we define as the ratio of data written to flash by the device to data the host system requested written and requires additional flash transactions.

\subsection{Performance Implications}

\subsubsection{Concurrency \& Throughput}
The hierarchical structure of modern flash memory is designed to maximize the potential for concurrent operations.
To facilitate this, the FTL maintains internal transaction queues for each chip while the NVMe driver bypasses the OS I/O stack, providing multiple submission queues directly to applications \cite{nvme, mqsim}. 
The device's ability to saturate these internal transaction queues significantly impacts the overall throughput of the device.  
Saturating these transaction queues generally requires high concurrency from applications.

\subsubsection{Asymmetric Reads \& Writes}
The primary source of latency for any I/O operation is the number of flash transactions required; however, not all transactions are equal. 
The physical operation of writing, or programming, a flash page has roughly 10x the latency of a read operation, and erasing an entire block is even slower \cite{mqsim}.  
Due to out-of-place updates, writes also have more programmatic steps, with each page update requiring an entire read operation before the contents are written to a new page.  
If a write triggers garbage collection, additional transactions will be required to free up more space.  
As an SSD is used, it will eventually reach a \textit{steady state} where the number of valid, invalid, and free pages stays pretty consistent.  
This state is reached once each physical page has been written at least once \cite{mqsim}.
Due to out-of-place updates and wear leveling, these writes could all be to a small portion of the logical address space, but the data will still be moved around the device to each physical page.   
The precise distribution of pages in a steady state depends on the amount of valid data on the device, the workload, and the garbage collection algorithm; however, most steady states have very few free flash pages \cite{steady_state}.
This means after a certain amount of use, a significant portion of writes to the device will trigger garbage collection, resulting in additional read, write, and erase transactions that will take up bandwidth. 
Therefore, not only are writes inherently more expensive, but they can also have a more significant impact on overall throughput.

\subsubsection{Address Translation \& Caching}
Due to out-of-place updates, the FTL must keep track of the mappings between the logical pages accessed by applications and the physical pages on the device.
These mappings are stored in a table on flash and cached in the FTL's DRAM.  
This DRAM is also used for internal transaction scheduling and write caching, so the amount of space for the cached mapping table is limited, and only recently accessed mappings can be assumed to be cached \cite{mqsim}. 
If a given mapping is not cached, it must be read from flash before the target page can be read. 
Every time a page is updated and moved to a new physical page, the mapping table must also be updated, requiring another transaction with flash.  
As with all other flash transactions, reads from and updates to the mapping table are done at page granularity, so $\Theta(B)$ logically sequential page mappings are read into or written back from the cache per transaction.  
Therefore, the logical locality of page accesses substantially impacts the efficiency of address translation. 

Consider if an application requests to read $N$ records stored sequentially.  
This results in a request for $\Theta(\frac{N}{B})$ logical pages, as each page is fully utilized, and the FTL only needs to read $\Theta(\frac{N}{B^2})$ pages from the mapping table as it retrieves $\Theta(B)$ useful mappings with each transaction.  
Conversely, if an application requests to read $N$ records stored on different pages at random logical addresses, not only will this result in a request for $\Theta(N)$ pages of records, since each page only contains a single record of interest, but the FTL will also need to read $\Theta(N)$ pages from the mapping table as each page is only guaranteed to have a single helpful mapping. 
Writes interact with the mapping table similarly, except they must also make updates.  
Both the initial reads and subsequent updates from and to the mapping table will be batched most efficiently if the logical page addresses have high locality, while writes with poor logical locality can require up to one mapping table read and update per page access.  
The resulting relationship between sequential and random I/O is similar to what we see in HDDs but with a few caveats.  
Whether accessed sequentially or not, recently accessed pages will not require a setup cost as the mapping is likely to be cached.
Also, due to the differences in operation, the costs for reads and writes are significantly different.

\subsection{Validation}
\label{sec:validation}
We now use a file I/O benchmark to demonstrate the performance characteristics we have just described.
These are similar in spirit to those used by Bender et al. to validate the PDAM model for SSDs \cite{refinements}, but we also consider writes and sequential access in addition to random reads.

\subsubsection{Experimental Setup}
We are interested in observing the entire performance spectrum concerning the order of access and concurrency.  
For a given $k$ and $r$, the benchmark will consist of $k$ threads concurrently reading or writing 1 GB of data per thread, each with $r$ random accesses to or from a 50 GB file.  
We get $r$ random accesses per thread by dividing the 1 GB into $r$ equally sized write or read operations starting at uniformly random offsets within the file. 
Using staggered offsets for each thread, we then vary $r$ from $0.1\%$ to $100\%$ of pages accessed and $k$ from $1$ to $128$.  

All experiments were run on an AMD Ryzen\texttrademark{} 5 7600X CPU @ 4.7 GHz x 6 with 32 GB of DDR5 RAM running Ubuntu Server with Linux kernel 5.15.0-97-generic.  
We use three different SSDs, which are described in Table \ref{tab:devices}, all of which are connected directly to the motherboard using M.2 PCIe$^\circledR$ 5.0 (which is backward compatible with previous PCIe$^\circledR$ generations). 
The SSDs are preconditioned by writing 2 \texttt{TB} (twice their capacity) of data to ensure that garbage collection will be active during the benchmarks, as per SSD evaluation standards \cite{snia}.  

\begin{table}
    \centering
    \begin{tabular}{|c|c|c|c|c|} \hline
        Mfr.  & Model & Firmware & NVMe & PCIe \\ \hline \hline
        Crucial & CT1000T700SSD5 & PACR5101 & 2.0 & 5.0 \\ \hline
        Samsung & SSD 990 PRO 1TB & 3B2QJXD7 & 2.0 & 4.0 \\ \hline
        PNY & CS1030 1TB SSD & CS103F00 & 1.3 & 3.0 \\ \hline
    \end{tabular} \vspace{+10pt}
    \caption{Multi-queue SSDs used for benchmarking.  }
    \label{tab:devices}
\end{table}

\subsubsection{Results}

\begin{figure*}
    \centering 
    \includegraphics[width=\textwidth]{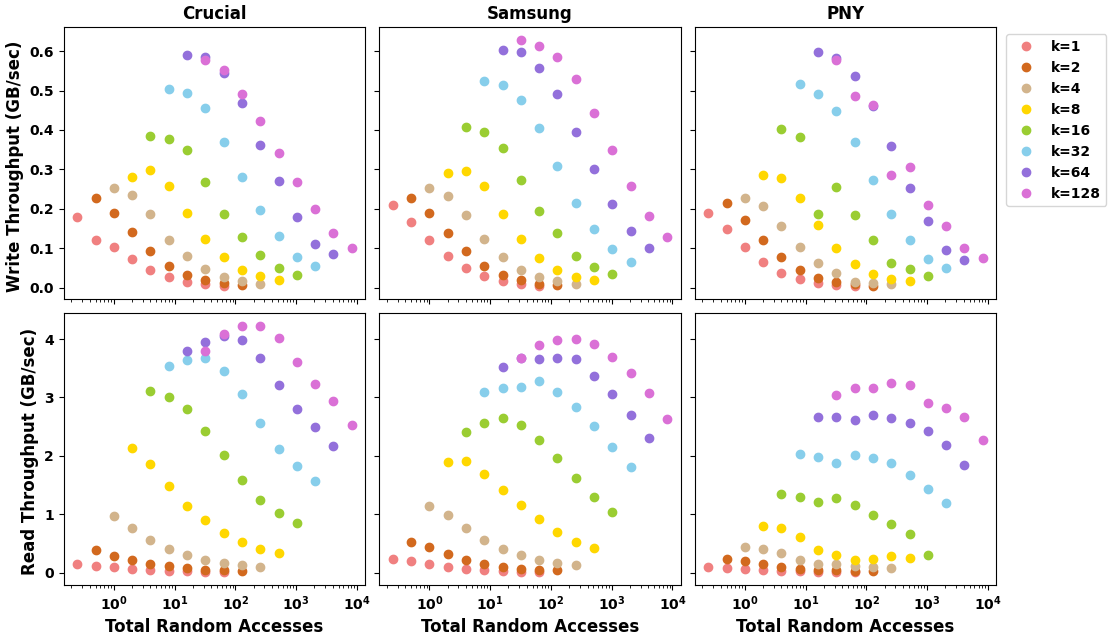}
    \caption{File I/O throughput with respect to concurrency and random access for devices in Table \ref{tab:devices}.  Performance varies significantly with respect to $r$ and $k$, but the trend is fairly consistent across all three devices, with some small variations, regardless of NVMe version or PCIe$^\circledR$ generation.  }
    \label{fig:microbench}
\end{figure*}

The results of the file I/O benchmark for all three devices are shown in Figure \ref{fig:microbench}.  
For all devices, we see a significant difference in cost between both reads and writes, resulting in a much higher throughput for reads than writes for any given values of $r$ and $k$. 
Despite different manufacturers, NVMe versions, and PCIe generations, all devices exhibit similar performance trends concerning $r$ and $k$. 
This comes down to the devices' abilities to saturate their internal transaction queues and the effects of address translation.  
The proportion of random page accesses substantially impacts throughput for all levels of concurrency, but the overall impact of $r$ decreases as $k$ increases.  
Read and write throughput scale differently with respect to $k$ and $r$, with reads being less affected by random access and gaining more benefit from concurrency.  
Single-threaded write throughput drops by a factor of $38 - 57x$ across all three devices as $r$ reaches $100\%$ of pages written; however, as $k$ increases, the drop in throughput with respect to $r$ drops to a factor of $5.0 - 7.8x$ across all devices.  
Single-threaded read throughput drops by $18 - 23x$ across all three devices as $r$ reaches $100\%$ of pages read.
This drops to $1.3 - 1.5 x$ across all devices as $k$ increases.  

The observed trends align with the performance bottlenecks described in Section \ref{sec:background:mqssds}.  
The extra steps and more expensive flash transactions associated with writes result in write operations taking longer and using up more of the device's bandwidth.  
Random access results in less efficient address translations and updates, resulting in additional flash transactions per page access for both reads and writes.  

\section{The Multi-Queue SSD (MQSSD) Model}
\label{sec:the_model}
We now compare previous models to observed performance and define a further refinement of the DAM model designed to capture the performance characteristics of multi-queue SSDs.
In particular, we focus on the effects of concurrency and order of access.  

\subsection{Lessons From Previous Models}

\begin{figure*}
    \centering
    \includegraphics[width=\textwidth]{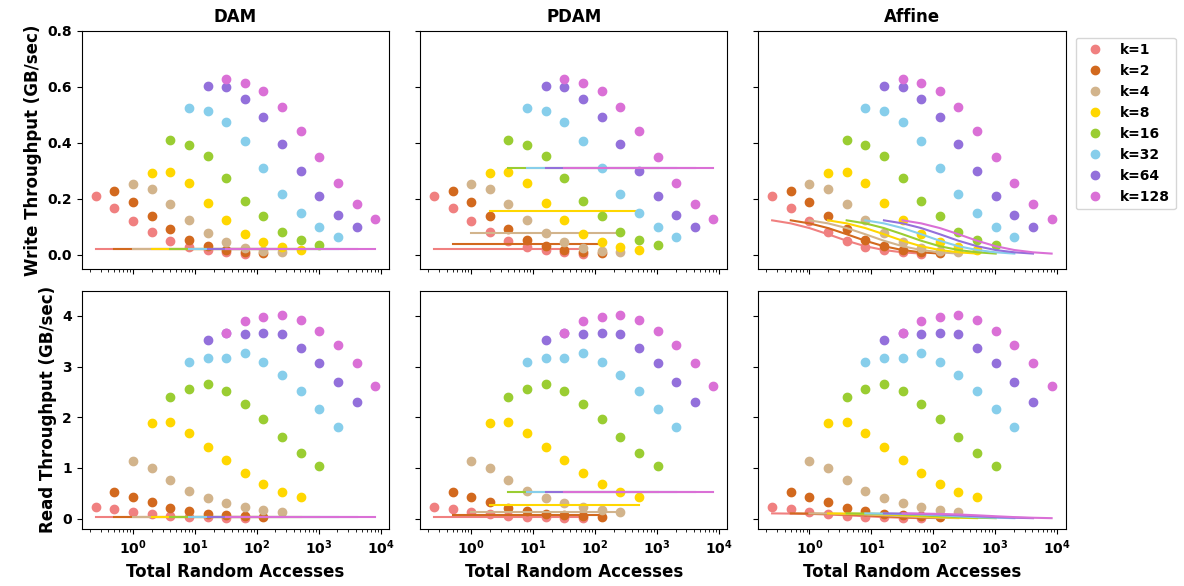}
    \caption{File I/O throughput predictions for the Samsung SSD 990 Pro using DAM, PDAM, and Affine models.  Each solid line is the model's prediction for the corresponding value of $k$, with $r$ ranging from $0.1-100\%$ of the pages read per thread.  The DAM model predicts constant throughput for all values of $r$ and $k$.  The PDAM model roughly captures how throughput scales with $k$ but significantly over and underestimates for writes and reads, respectively, as $k$ increases and has no consideration for $r$.  The Affine model captures how throughput scales with $r$ but has no consideration for $k$.}
    \label{fig:model_comp}
\end{figure*}

Despite being based on HDDs, aspects of each of the previous models we have discussed are still applicable to multi-queue SSDs.
To demonstrate their proficiencies and shortcomings, we consider how each model would represent the file I/O benchmark from Section \ref{sec:background:mqssds} and compare them to the results.
These comparisons are visualized in Figure \ref{fig:model_comp} using the Samsung SSD as the reference device.  
As discussed, none of these models distinguish between the costs of reads and writes since both operations function similarly in HDDs.  
To give these models the best chance at an accurate prediction, the predictions in Figure \ref{fig:model_comp} use different values for the DAM model's \texttt{page\_cost}, the PDAM model's \texttt{cycle\_cost}, and the Affine model's setup ($s$) and page transfer ($\beta$) costs for reads vs. writes.
These device-specific values are derived for the model based on the single-threaded benchmarking results. 
For \texttt{page\_cost} and \texttt{cycle\_cost}, we use the average time per page access, while for $s$ and $\beta$, we use a least-squares linear regression.

\noindent \textbf{DAM Model:} 
Since the DAM model does not consider the order of access or concurrency, it would suggest that all trials of our benchmark will have the same throughput.  
If we take $N$ to be the amount of data written per thread, Equation \ref{eq:dam_cost} gives the predicted throughput in the DAM model.
\begin{equation}
    \label{eq:dam_cost}
    \texttt{Throughput}_{\texttt{DAM}}(r, k) = \frac{B}{\texttt{page\_cost}}  
\end{equation}
As shown in Figure \ref{fig:model_comp}, Equation \ref{eq:dam_cost} results in flat overlapping lines since it assumes only a single thread can access the device at a time and considers all pages to be of the same cost.
While this model still accurately captures the total amount of I/O for a given workload, it fails to account for how I/O costs change concerning $r$ and $k$, even if we adjust the cost for reads vs. writes.

\noindent \textbf{PDAM Model:}
The PDAM model's throughput predictions are very similar to the DAM model's, but with some consideration for concurrent access, represented by Equation \ref{eq:pdam_cost}.
\begin{equation}
    \label{eq:pdam_cost}
    \texttt{Throughput}_{\texttt{PDAM}}(r, k) = \begin{cases} 
            \frac{ B * k }{\texttt{cycle\_cost}}, & k < P \\
            \frac{ B * P }{\texttt{cycle\_cost}}, &  k \geq P\end{cases}
\end{equation}
The predictions are still flat lines with respect to $r$ since all pages are considered equal, but the throughput scales with $k$ up to $P$.  
This model generally has the same shortcomings as the DAM model, except that it has a basic abstraction for concurrency.  
Despite this, the PDAM model cannot capture how concurrency affects reads and writes differently, resulting in significant under and overestimates in throughput as concurrency increases.

\noindent \textbf{Affine Model:}
In the Affine model, we again have no consideration for concurrency, but we take into account the order of access.  
The throughput is then determined by the page transfer cost, $\beta$, the setup cost, $s$, and the proportion of random accesses, $\frac{r * B}{N}$, as given by Equation \ref{eq:affine_cost}.
\begin{equation}
    \label{eq:affine_cost}
    \texttt{Throughput}_\texttt{Affine}(r, k) = \frac{N}{s * r} +\frac{B}{\beta}
\end{equation}
This model accurately reflects how single-threaded performance changes with rest to $r$, but it is unable to account for how this changes with respect to concurrency. 
While Figure \ref{fig:model_comp} uses $s$ and $\beta$ derived from the single-threaded performance, we could just as easily pick any other value of $k$, which results in a good fit with respect to $r$ for only that value of $k$.

\subsection{Definition}
\label{sec:definition}
The Multi-Queue SSD (MQSSD) model builds upon the DAM model by applying insights from both the PDAM and Affine models to multi-queue SSDs.   
The parameters for the MQSSD model are defined as follows:
\begin{itemize}
    \item[$N$:] The size of our working set.  For our analysis, we will use bytes.  
    \item[$B$:] The page size of the device (typically 4096 bytes). 
    \item[$M$:] The size of main memory. At most $\frac{M}{B}$ pages can reside in memory simultaneously. 
    \item[$k$:] The number of threads/processes concurrently accessing the device.  
    \item[$s(k)$:] The setup cost for writes as a function of concurrent accesses, i.e., the time required to translate and schedule $k$ concurrent writes without cached mappings (random access).
    \item[$\beta(k)$:] The page transfer cost for writes as a function of concurrent accesses, i.e., the time required to write $k$ pages with cached mappings concurrently.  This is the only cost for sequential access but is additive with $s(k)$ for random access.
    \item[$t(k)$:] Analogous to $s(k)$ but for reads.  
    \item[$\alpha(k)$:] Analogous to $\beta(k)$ but for reads.  
\end{itemize}

The MQSSD model uses the same basic framework of page transfers to and from memory as previous models, with $N$, $B$, and $M$ defined as in the DAM model.  
We also use separate setup and page transfer costs, as in the Affine model, though in this context, the setup cost is due to address translations and mapping updates rather than seek time.
In addition to this, we distinguish between the setup and page transfer costs for writes ($s(k)$ \& $\beta(k)$) vs. those for reads ($t(k)$ \& $\alpha(k)$). 
Finally, rather than using fixed costs, as in the affine model, our setup and page transfer costs are functions of concurrency.  
This can be considered a generalization of how the PDAM model accounts for concurrency.  
For example, if we use Equation \ref{eq:gen_pdam} as the \texttt{page\_cost} in the DAM model, this would be equivalent to the PDAM model.  
\begin{equation}
    \texttt{page\_cost}_\texttt{PDAM}(k) = \begin{cases}
        \frac{\texttt{cycle\_cost}}{k}, & k < P \\
        \frac{\texttt{cycle\_cost}}{P}, & k \geq P
    \end{cases}
    \label{eq:gen_pdam}
\end{equation}

\begin{figure}
  \begin{minipage}[b]{.7\linewidth}
    \centering
    \begin{tabular}{|c|c|c|c|c|c|c|} \hline
        \multicolumn{7}{|c|}{Derived Samsung SSD Costs for Each $k$ ($\mu$sec)} \\ \hline
        $k$ & $s(k)$ & $\beta(k)$ & $R^2$ & $t(k)$ & $\alpha(k)$ & $R^2$ \\ \hline \hline
        1 & 3.10x10$^5$ & 2.79 & 0.996 & 1.47x10$^5$ & 3.53 & 0.969 \\ \hline
        2 & 2.54x10$^5$ & 3.01 & 0.989 & 4.07x10$^4$ & 1.71 & 0.909 \\ \hline
        4 & 1.54x10$^5$ & 2.59 & 0.981 & 1.05x10$^4$ & 0.659 & 0.885 \\ \hline
        8 & 8.02x10$^4$ & 1.93 & 0.971 & 2.93x10$^3$ & 0.265 & 0.913 \\ \hline
        16 & 4.23x10$^4$ & 1.27 & 0.976 & 931 & 0.157 & 0.957 \\ \hline
        32 & 2.15x10$^4$ & 0.891 & 0.981 & 392 & 0.121 & 0.973 \\ \hline
        64 & 1.34x10$^4$ & 0.745 & 0.975 & 263 & 0.105 & 0.979 \\ \hline
        128 & 9.98x10$^3$ & 0.697 & 0.977 & 202 & 0.0974 & 0.964 \\ \hline

    \end{tabular} \vspace{+10pt}
    \captionof{table}{Setup costs ($s(k),t(k)$), page transfer costs ($\beta(k),\alpha(k)$), and the respective $R^2$ values for the Samsung SSD 990 Pro derived for each thread count ($k$) using a least squares linear regression on the benchmarking data shown in Figure \ref{fig:microbench}. }
    \label{tab:model_params}
  \end{minipage}\hfill
  \begin{minipage}[b]{.25\linewidth}
    \centering
    \includegraphics[width=\linewidth]{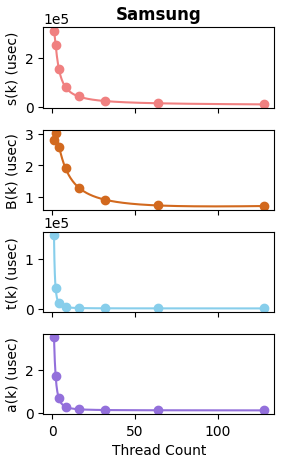}
    \captionof{figure}{Derived cost functions for the Samsung SSD 990 Pro. }
    \label{fig:model_params}
  \end{minipage}
\end{figure}

To see how the setup and page transfer costs change with respect to concurrency, we use the same process as with the Affine model but for all values of $k$.  
We use a least squares linear regression on the benchmarking data for each value of $k$ to derive the respective setup and page transfer costs for reads and writes, respectively.  
These derived values for the Samsung device are shown in Table \ref{tab:model_params} along with the corresponding $R^2$ values. 
Not only do the resulting setup and page transfer costs give us a reasonably strong fit for any particular value of $k$, but they also demonstrate the substantial difference in performance between sequential and random I/O.  
We observe that the setup and page transfer costs drop very quickly for small $k$ as $k$ increases and continue to drop for large $k$ but with diminishing returns.  
While we see the same trend for both reads and writes, the read costs diminish much more quickly with increasing $k$ since they require fewer and cheaper flash transactions than writes and take up less bandwidth.  

We can capture this type of relationship using rational functions.  
In particular, we can fit our setup and page transfer costs using Equations \ref{eq:param_cost} and \ref{eq:write_page} where the $c_i$ and $d_i$ are constants that will vary for each function depending on the particular hardware, similar to $s$ and $\beta$ in the Affine model.
Figure \ref{fig:model_params} shows the setup and page transfer cost functions overlaying the derived values from Table \ref{tab:model_params}.
These functions were fit to Equations \ref{eq:param_cost} and \ref{eq:write_page} using the \texttt{scipy.optimize} library \cite{scipy}.

\noindent\begin{minipage}{.5\linewidth}\vspace{+3pt}
  \begin{equation}
    \label{eq:param_cost}
    t(k), \alpha(k), s(k) = \frac{c_0 + c_1 k + c_2 k^2}{d_0 + d_1 k + d_2 k^2}
  \end{equation}\vspace{+3pt}
\end{minipage}%
\begin{minipage}{.5\linewidth}\vspace{+3pt}
  \begin{equation}
    \label{eq:write_page}
    \beta (k) = \frac{c_0 + c_1 k + c_2 k^2 + c_3 k^3}{d_0 + d_1 k + d_2 k^2 + d_3k^3}
  \end{equation}\vspace{+3pt}
\end{minipage}

The general idea is that the device has some maximum bandwidth for flash transactions, and as we add more concurrency, we can utilize more of that bandwidth.  
However, due to the complexities of flash memory (hierarchical structure, out-of-place updates, transaction scheduling, garbage collection, etc.), fully utilizing the device's potential bandwidth is non-trivial.
For example, the order and type of transactions scheduled for a particular chip will determine how effectively it can distribute commands over each die.  
Some transactions will also have to be scheduled for particular chips. 
Read transactions are limited by where the target data physically resides, while write transactions can only be assigned to free pages.   
Increasing the number of concurrent I/O operations results in more transactions that can be scheduled more efficiently across the various chips and dies; however, the complexity of this scheduling also increases with the number of transactions.  
The cost per page or random access continues to diminish even for large values of $k$ but will approach a constant as we approach the device's maximum potential bandwidth.  
This relationship is reflected by Equations \ref{eq:param_cost} and \ref{eq:write_page}, which approach a constant for large $k$.
We use these rational functions to capture the nuances of how performance improves for low $k$ \footnote{We tested rational functions with lower degrees, but these resulted in inconsistent spikes in our cost functions.  The degrees shown were the minimum required to obtain the smooth curves shown in Figure \ref{fig:model_params}.}, but these can ultimately be simplified for analysis. 
For Equation \ref{eq:write_page} for $\beta(k)$, we use a higher degree numerator and denominator to capture the additional complexities associated with garbage collection. 
In general, more concurrency will almost always improve I/O throughput on these devices, but the impact will eventually start to diminish. 
For writes, in particular, the amount of random access significantly impacts how well throughput scales with $k$. 
As a result, data structures that facilitate concurrent and sequential I/O are likely to achieve the best throughput.

\subsection{Validation}

\begin{figure*}
    \centering
    \includegraphics[width=\textwidth]{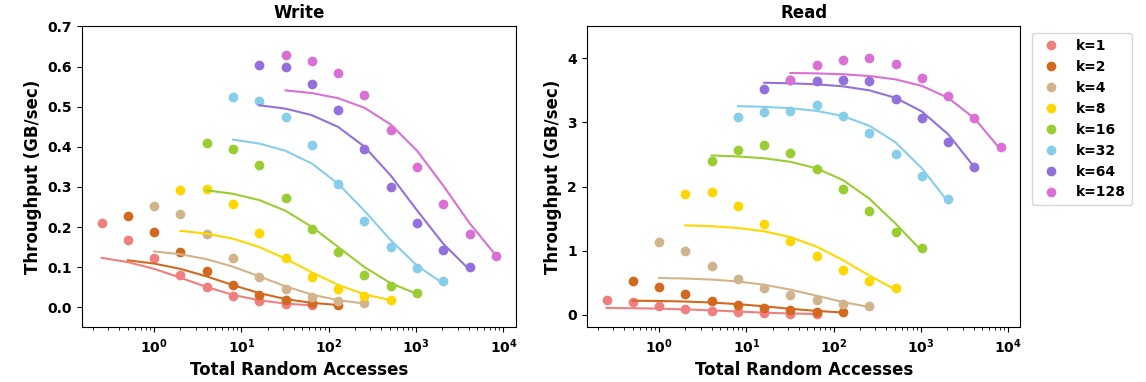}
    \caption{File I/O throughput predictions for the Samsung SSD 990 Pro using the MQSSD model.  Each solid line is the model's prediction for the corresponding value of $k$, with $r$ ranging from $0.1-100\%$ of the pages accessed per thread.  The MQSSD model captures the effects of $r$ and $k$, how they affect each other, and the differences between reads and writes. }
    \label{fig:mqssd}
\end{figure*}

Returning to the file I/O benchmark from earlier, we compare the MQSSD model's predictions to the benchmarking results for the Samsung SSD 990 Pro. 
In the MQSSD model, we can represent the throughput for this benchmark using Equations \ref{eq:mqssd_write_cost} and \ref{eq:mqssd_read_cost} for writes and reads, respectively.
These are essentially the same as the cost for the Affine model, except that our setup and page transfer costs distinguish between reads and writes and change with respect to $k$, as per Equations \ref{eq:param_cost} and \ref{eq:write_page} and Figure \ref{fig:model_params}. 
These predictions using the fitted cost setup and page transfer cost functions are shown in Figure \ref{fig:mqssd}, overlaying the benchmarking results.  

\noindent\begin{minipage}{.5\linewidth}\vspace{+3pt} 
  \begin{equation}
    \label{eq:mqssd_write_cost}
    \texttt{W\_Throughput}_\texttt{MQSSD}(r, k) = \frac{N}{s(k) * r} +\frac{B}{\beta(k)} 
  \end{equation}\vspace{+3pt} 
\end{minipage}%
\begin{minipage}{.5\linewidth}\vspace{+3pt} 
  \begin{equation}
    \label{eq:mqssd_read_cost}
    \texttt{R\_Throughput}_\texttt{MQSSD}(r, k) = \frac{N}{t(k) * r} +\frac{B}{\alpha(k)} 
\end{equation}\vspace{+3pt} 
\end{minipage} 

The MQSSD model's predictions do not perfectly match the observed performance, but they do capture the effects of random access and concurrency far more accurately than any of the previous models.  
In particular, it is able to predict how throughput degrades as $r$ increases and how this scales with respect to $k$. 
How we organize data layouts for storage directly impacts the amount of sequential vs. random I/O required for various operations and the amount of concurrency that can be utilized.  
The MQSSD model provides a formal framework for analyzing the impact of these design choices.

\section{Applications of the Model}
\label{sec:applications}
Even for an engineer familiar with the code base and the implications of various configuration parameters, tuning a database for a new workload or hardware can involve significant trial and error, which costs time and resources. 
Here, we demonstrate how the MQSSD model can be used to understand the performance implications of various tuning parameters for one of the most common data structures used as the storage layer of modern NoSQL key-value stores, the Log-Structured Merge (LSM) tree, when backed by a multi-queue SSD \cite{lsm_survey, key-value_stores}. 
LSM-trees have been deployed in several industrial settings, including LevelDB \cite{leveldb}, Facebook's RocksDB \cite{rocksdb}, Google's Bigtable \cite{bigtable}, Apache HBase \cite{hbase}, LinkedIn's Voldemort \cite{voldemort}, Amazon's Dynamo \cite{Dynamo} as well as Yahoo's bLSM \cite{blsm} and cLSM \cite{clsm}. 
We use RocksDB as our reference architecture for analysis and benchmarking.  
As reported by Facebook \cite{space_rocks}, RocksDB has been used in several production settings, including real-time data processing \cite{realtime_rocks}, graph processing \cite{fb_dragon}, stream processing \cite{realtime_rocks}, and online transaction processing (OLTP) \cite{myrocks}.

\subsection{LSM Trees \& Compaction}
Unlike a B-tree, LSM trees forgo global ordering and in-place updates to support faster ingestion.
Data is partitioned temporally into levels of exponentially increasing size. 
Incoming key-value pairs are first buffered in memory and then flushed to storage as sorted immutable runs, which are later sort-merged using sequential I/O. 
Runs are first flushed to the smallest level, $L_0$, and as data accumulates in a given level, it will eventually be sort-merged into the next level \cite{the_lsm}. 
This sort-merge operation, referred to as \textit{compaction}, is fundamental to an LSM-tree's performance as it reduces read and space amplification at the cost of write amplification\footnote{In this section, write amplification refers to data structure write amplification as defined in Section \ref{sec:backgroud:ex}.  Read and space amplification are defined analogously. } \cite{rocksdb, lsm_survey, compaction_design}.  
Additional strategies are used to reduce I/O and improve read performance, including auxiliary indexing and membership structures like fence pointers, tries, and probabilistic filters \cite{monkey, rocksdb, proteus, surf}.

Sarkar et al. present the compaction design space as an ensemble of 4 primitives: (\textit{i}) the \textit{trigger} or \textbf{when} we decide to compact two or more runs, (\textit{ii}) the \textit{data layout} or \textbf{how} the data is organized, (\textit{iii}) the \textit{granularity} or \textbf{how much} data is compacted at once, and (\textit{iv}) the \textit{data movement policy} or \textbf{which} data should be compacted \cite{compaction_design}. 
We recommend their work for a more detailed description of the compaction design space, including an online compaction dictionary, \href{https://disc-projects.bu.edu/compactionary/index.html}{Compactionary} \cite{compactionary}, with helpful visualizations.  
For this demonstration, we focus on the I/O performance implications of the data layout when backed by a multi-queue SSD. 
Below, we describe how RocksDB implements these primitives and define the terms for our analysis.

\subsubsection{Compaction in RocksDB}
In RocksDB, all sorted runs are partitioned by key range into Static Sorted Table (SST) files of roughly the same size.
The first level, $L_0$, uses a \textit{tiered} layout, while all subsequent levels used a \textit{leveled} layout.  
This means that each file in $L_0$ is considered a separate sorted run and may have overlapping key ranges. 
As with tiering in general, this reduces write amplification at $L_0$, as no data is ever rewritten within the level, but it incurs higher read and space amplification since the files have overlapping key ranges and may contain outdated values.  
By contrast, leveling maintains a single sorted run per level, which results in lower read and space amplification at the cost of higher write amplification since we perform a sort-merge every time more data is added to the level \cite{dostoevsky, lsm_survey, lsm, compaction_design}.  
The standard compaction trigger is when any level exceeds its capacity, as determined by the capacity growth factor between levels.  
The size of $L_0$ is controlled independently.
The RocksDB documentation suggests making $L_0$ and $L_1$ similar in size, which is how we have configured our benchmarks \cite{rocksdb}.
Once compaction is triggered at $L_i$, the data movement policy picks the file in $L_i$ with a key range that overlaps with the fewest files in $L_{i+1}$ to minimize write amplification.  
The selected file and all its overlapping files in $L_{i+1}$ are merged and rewritten to $L_{i+1}$.
Multiple files can be chosen to return $L_i$ to its target size if required.  
Multiple compactions can be performed in parallel, provided they do not touch any of the same files.  
Because the files in $L_0$ likely have overlapping key ranges, most compaction operations from $L_0$ to $L_1$ would need to touch some of the same files, severely limiting the potential for concurrency.  
To circumvent this, RocksDB implements a special type of compaction called subcompactions, specifically for $L_0$ to $L_1$. 
Subcompactions partition individual compactions by key range into multiple concurrent operations, allowing for up to as many concurrent compactions as there are files in $L_1$.

\subsubsection{Problem Definition}
Consider a data set, $S$, consisting of $N$ bytes of key-value pairs with a uniform distribution of keys.  
We want to compare the performance of inserts, updates, deletes, queries, and scans for RocksDB's LSM tree built on $S$ as we tune its data layout. 
In an LSM tree, inserts, updates, and deletes are all executed similarly: a timestamped key-value pair is inserted, either with a new or existing key. 
If the key already exists in the database, the key with the most recent timestamp is the actual value. 
Matching keys are combined into a single key-value pair during compaction.  
When compacting into the last level, the key is removed entirely if the most recent value is a delete message.  
Our measure of performance for these operations is the compaction throughput.
For queries, we consider the average time required to retrieve a single key-value pair, whereas for scans or range queries, we consider the time needed to retrieve all keys within a given range.   
As mentioned earlier, auxiliary structures are also used to improve query performance.  
These structures are maintained per SST file level, with each SST file containing indexing and other information for each data block.  
The default size for the data blocks is 4 kilobytes, but this can be modified.  
Pinning the metadata blocks in memory is common practice to minimize the amount of I/O necessary for queries, assuming there is sufficient space \cite{key-value_stores, lsm_survey, monkey}.  
For example, one of the indexing structures RocksDB stores in these meta blocks are fence pointers, which provide the min and max keys for each data block.  
With these fence pointers pinned in memory, RocksDB can determine which data blocks could contain a target key without I/O and then read only the necessary data blocks rather than the whole file or run.
With Bloom filters or other approximate membership structures, much of this I/O can be avoided entirely, particularly for point queries \cite{monkey, proteus}.  
We will assume that fence pointers are pinned in memory for our analysis.  
Other parameters of the data layout that will be relevant are: 
\begin{itemize}
    \item[$F$:] Fanout, or the capacity growth factor between levels. 
    \item[$T$:] Target SST file size, or the maximum granularity for sequential accesses.  
    \item[$C$:] $L_0$ file count.  
    \item[$B'$:] DB block size, i.e., the smallest granularity used by the DB for I/O.
\end{itemize}

\subsection{RocksDB in the MQSSD Model}
Before discussing the details of compaction, we consider the overall shape of the structure and the order of access for basic operations.    
For $i > 0$, level $i$, or $L_i$, will have a capacity of $TC\times F^{i-1}$ bytes, for a total of $O \left ( \log_{F} \frac{N}{CT} \right )$ levels and $O\left (C + \log_{F} \frac{N}{CT} \right )$ sorted runs since we have $C$ files in $L_0$.   
Reading and rewriting $X$ files would require $\frac{XT}{B}$ page accesses, $X$ of which will be random, one per SST file.  
Using $k$ concurrent threads, this would have a cost of $X\left(t(k) + s(k) + \frac{T\left(\alpha(k) + \beta(k)\right)}{B}\right)$. 
Assuming fence pointers are pinned in the block cache, queries will access at most one data block per sorted run, while range queries will only access the data blocks overlapping with the target key range in each sorted run.  

\subsubsection{Modeling Compaction}
We now consider the cost of compacting a file from $L_i$ into $L_{i+1}$. 
As described earlier, the file chosen for compaction is the one that touches the least files in $L_{i+1}$. 
For the case of $i=0$, we will almost always have to touch all $C$ files at $L_1$ since $L_0$ is tiered and all files can cover the entire key range. 
For $i > 0$, the expected minimum number of files touched is related to $F$ but also depends on the distribution of keys and the order in which they are inserted. 
This is particularly true for RocksDB's deamortized compactions and data movement policy, which aim to minimize the number of files touched per compaction.  
For our benchmarking, we use RocksDB's built-in \texttt{db\_bench} toolkit, so we know the key set consists of the keys $0$ to $N-1$ and are inserted uniformly at random.
Based on this, we can simulate RocksDB's data movement policy for our workload to derive the expected number of files touched per compaction as a function of $F$.
For this particular workload, the expected number of files touched empirically appears to be $\Theta(F)$, with a least squares linear regression resulting in an $R^2$ of $0.974$.
This aligns with the standard LSM tree analysis, though this will not be true for all workloads.

Another consequence of RocksDB's deamortized compaction is that all levels except the last will always be at or near capacity.  
This means that each time we add a new file to the data structure, cascading compactions are likely to occur at every level. 
This then touches $O(C)$ files at $L_1$ and $O(F)$ at each of the $O\left(\log_F\frac{N}{CT}\right)$ subsequent levels. 
However, since we only add to the on-storage structure in batches of $O(T)$, we amortize the per key-value cost of an insert, update, or delete, as shown by Equation \ref{eq:insert}.
\begin{equation} \label{eq:insert}
\begin{split}
    I\left(F, D, k\right)  &= O\left(\frac{C + F \log_{F} \frac{N}{CT}}{T} \left(t(k) + s(k) + \frac{T\left(\alpha(k) + \beta(k)\right)}{B}\right)\right)
\end{split}
\end{equation}

We now consider the cost of queries and scans.
Assuming fence pointers are pinned in memory, a point query requires one data block to be read per sorted run, each of which has a cost of $t(k) + \frac{B'\alpha(k)}{B}$.
Equation \ref{eq:query} then gives us the expected point query cost as we have $O\left(C + \log_F\frac{N}{CT}\right)$ sorted runs.  
\begin{equation} \label{eq:query}
\begin{split}
    Q\left(F, C, k\right)  =  O\left(\left(C + \log_{F} \frac{N}{CT}\right) \left(t(k) + \frac{B'\alpha(k)}{B}\right)\right)               
\end{split}
\end{equation}
A scan retrieving $X$ bytes of data will then have a cost equivalent to the point query cost with an additional cost of $O\left( X \left(\frac{t(k)}{T} + \frac{\alpha(k)}{B}\right)\right) $ to scan through the files containing the data.

\subsubsection{Insights}
\label{sec:insights}
As with the DAM model, our analysis of leveled compaction in RocksDB gives us the standard tradeoff between read and write performance concerning $F$.
A larger fanout results in more write amplification due to more aggressive merging and less read amplification since there are fewer total sorted runs.  
However, RocksDB's deamortized compaction and data movement policy reduces the amount of write amplification per level.
As with the Affine model, we see a similar tradeoff between write amplification and scan performance concerning $T$. 
Larger files mean we must rewrite more data during each compaction, but they also result in more sequential data and fewer random accesses when performing scans and compactions.  
However, this tradeoff is slightly more nuanced, as larger files only increase the number of page transfers during compaction.
To achieve the best throughput, files should be large enough to amortize the cost of random access.
Finally, all of our costs are directly affected by the level of concurrency that the data structure can facilitate.  
As shown in Table \ref{tab:model_params} and Figure \ref{fig:model_params}, the level of concurrency can reduce our setup costs by as much as 30x and 700x and page transfer costs by 4x and 35x for writes and reads, respectively.  
This suggests that facilitating concurrency has the potential to outweigh other tradeoffs.

The performance tradeoffs in an LSM tree concerning fannout and file size are both reasonably well understood \cite{compaction_design, key-value_stores}. 
RocksDB already strikes a good balance with respect to file size, using $64$ \texttt{MB} files by default, and offers some guidelines on how to pick $F$ based on your needs, with $F=10$ being the default \cite{rocksdb}.  
There has also been work scaling LSM trees for multi-core machines, but this has focused primarily on concurrency control for the in-memory portion of the structure and crash recovery \cite{clsm, lsm_survey}. 
Despite this, most LSM tree implementations do not fully utilize the capabilities of multi-queue SSDs, particularly concerning I/O concurrency \cite{nvme_perf_gap}. 
Looking at RocksDB specifically, its implementation can leverage as much read concurrency as the user provides \cite{key-value_stores, rocksdb} using out-of-place updates and asynchronous reads. 
However, its ability to facilitate concurrent compactions is more limited.
Multiple compactions can occur at different levels or within the same level if they do not touch the same files.  
The only direct control a user has over this concurrency is the number of threads available for compaction and the maximum number of subcompactions. 
By default, RocksDB only allocates a single background thread for compaction.
One can reduce the file size to facilitate more concurrency, but this reduces the performance of both scans and compactions.    
 
Based on these insights, we will also consider a somewhat unorthodox layout: a single level. 
In RocksDB, we can do this by setting $C=0$ and $F\geq N$. 
This takes the idea of leveling to its logical extreme; we maintain a single sorted run ($F=\infty$) by aggressively merging all incoming data into $L_1$.
This layout has two advantages: 1) its ability to leverage RocksDB's subcompactions mentioned earlier to achieve higher levels of concurrency, and 2) only having a single level achieves the best possible scan and query times with respect to $F$.   
Subcompactions can split a given compaction operation into as many files as they touch, but the size of $ L_1$ also limits the number of files touched. 
This layout is not typically used due to its very high write amplification. 
Each time we add a new file to $L_0$, it will be immediately compacted into $L_1$. 
Since files in $L_0$ can cover the entire key range, the number of files touched per compaction will approach $O(T)$ as $N$ grows.  
This means the write amplification and the concurrency level for compactions will approach the number of key-value pairs per file, $O(T)$. 
We can then replace the number of files touched in Equation \ref{eq:insert}, $\left(C+ F\log_{F} \frac{N}{CT} \right)$, with $T$ and we get Equation \ref{eq:sl_insert} for the per key-value pair compaction cost of this single level layout. 
Since there is only a single level and only one data block must be read per query, the cost of point queries will only depend on the number of concurrent threads, shown by Equation \ref{eq:sl_query}.
As before, scans will have the same cost as queries with an additional $O\left( X \left(\frac{t(k)}{T} + \frac{\alpha(k)}{B}\right)\right) $ to retrieve $X$ bytes of key-value pairs.

\noindent\begin{minipage}{.5\linewidth} \vspace{+3pt}
  \begin{equation} \label{eq:sl_insert}
    I_\texttt{SL}\left(k\right)  = O\left(t(k) + s(k) + \frac{T\left(\alpha(k) + \beta(k)\right)}{B}\right)
  \end{equation}\vspace{+3pt}
\end{minipage}%
\begin{minipage}{.5\linewidth}\vspace{+3pt}
  \begin{equation} \label{eq:sl_query}
    Q_\texttt{SL}\left(k\right)  =  O\left(t(k) + \frac{B'\alpha(k)}{B}\right)    
  \end{equation}\vspace{+3pt}
\end{minipage}

\subsection{Validation}

\subsubsection{Experimental Setup}
We validate our analysis against RocksDB v8.6.7 with its built-in benchmarking tool \texttt{db\_bench} \cite{rocksdb} on the same hardware described in Section \ref{sec:validation} using the Samsung SSD 990 PRO 1TB. 
We execute random inserts, point queries, and scans using the \texttt{filluniquerandom}, \texttt{readrandom}, and \texttt{seekrandom} benchmarks, respectively. 
The insert workload adds the keys $1$ to $6.4x10^7$ in random order with a single user thread.  
We control concurrency by adjusting the size of RocksDB's pool of compaction threads and the number of subcompactions allowed. 
We use \texttt{waitforcompaction} to ensure that all pending flushes and compactions run to completion.
The point query and scan benchmarks execute $3.2\times10^6$ random queries or scans with $k$ user threads on a database created using the insert workload. 
Each query is for a random key in the database, while scans read 128 key-value pairs starting at a random key.  
For both read workloads, the database is read sequentially using \texttt{readtocache} to ensure all metadata blocks are pinned in memory. 

For all workloads, we set the number of $L_1$ files to $C$, except for our single-level layout, which immediately merges any file added to $L_0$ into $L_1$ with no limit on the size of $L_1$.
All benchmarks use 16-byte integer keys with 2048-byte values for approximately $128$ \texttt{GB} of data. 
Direct I/O is turned on to best utilize the hardware, and compression is turned off to maintain consistency in storage accesses across experiments. 
Unless otherwise stated, all other settings use RocksDB's default values.  
All reported times are an average of 5 trials. 

\subsubsection{Results}

\begin{figure}
    \centering
    \includegraphics[width=\textwidth]{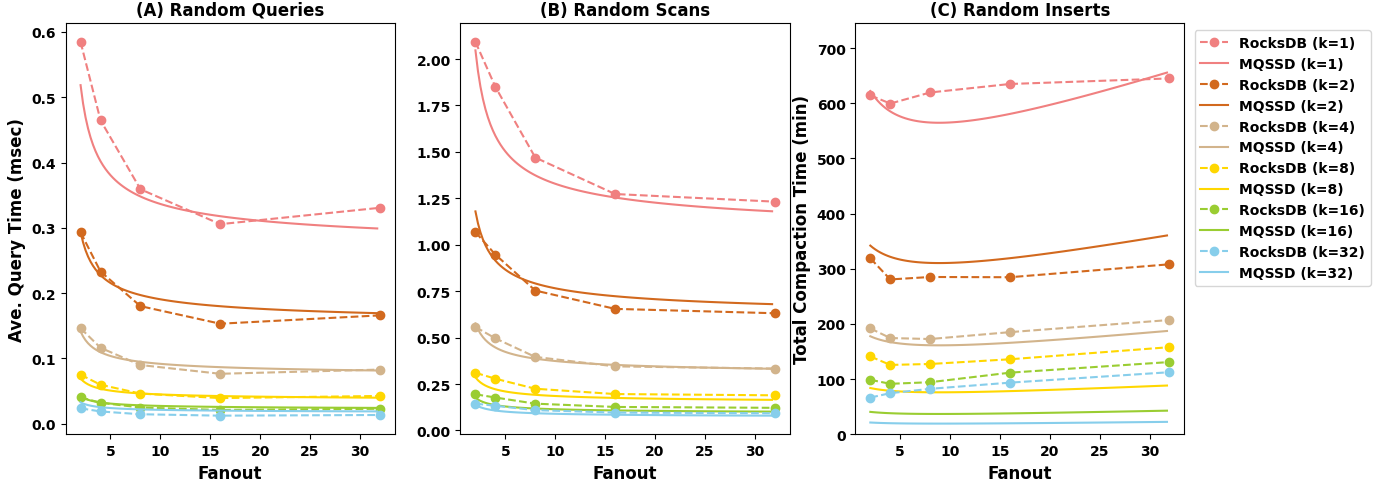}
    \caption{RocksDB performance for random queries, scans, and inserts with respect to fanout and concurrency and the MQSSD models predictions assuming full concurrency. The MQSSD model captures relative performance over both parameters, though RocksDB cannot achieve full concurrency for random inserts at high values of $k$.}
    \label{fig:rocksdb}
\end{figure}

\begin{table}
    \centering
    \begin{tabular}{|c||c|c|c|c|c|c||c|c|c|c|c|c|} \hline
         & \multicolumn{6}{c||}{Total Query Times (sec)} & \multicolumn{6}{c|}{Total Scan Times (sec)}\\ \hline
         & $F=2$ & $F=4$ & $F=8$ & $F=16$ & $F=32$ & S.L. & $F=2$ & $F=4$ & $F=8$ & $F=16$ & $F=32$ & S.L. \\ \hline \hline
        $k=1$ & 1870 & 1490 &  1150 & 978 & 1060 & 627 & 6700 & 5920 & 4700 & 4070 & 3940 & 3050\\ \hline
        $k=2$ & 938 & 745 & 576 & 490 & 531 & 314 & 3420 & 3030 & 2410 & 2100 & 2020 & 1570\\ \hline
        $k=4$ & 468 & 372 & 287 & 245 & 265 & 157 & 1780 & 1590 & 1270 & 1100 & 1070 & 835 \\ \hline
        $k=8$ & 240 & 191 & 148 & 125 & 136 & 80.6 & 994 & 892 & 719 & 628 & 606 & 481 \\ \hline
        $k=16$ & 129 & 102 & 79.1 & 67.2 & 73.0 & 43.2 & 624 & 568 & 462 & 406 & 391 & 317 \\ \hline
        $k=32$ & 75.2 & 59.8 & 46.2 & 39.3 & 42.6 & 25.2 & 462 & 425 & 350 & 308 & 297 & 249 \\ \hline

    \end{tabular} \vspace{+10pt}
    \caption{Total query and scan times for the \texttt{readrandom} and \texttt{seekrandom} benchmarks, respectively, for different numbers of user threads ($k$) and fanouts ($F$) as well as the single-level layout (S.L.). }
    \label{tab:query}
    \label{tab:scan}
    \vspace{-10pt}
\end{table}

\begin{table}
    \centering
    \begin{tabular}{|c||c|c|c|c|c|c|} \hline
        \multicolumn{7}{|c|}{Total Compaction Times (sec)} \\ \hline
          & $F=2$ & $F=4$ & $F=8$ & $F=16$ & $F=32$ & S.L. \\ \hline \hline
        $k=1$ & 3.69x10$^4$ & 3.60x10$^4$ & 3.72x10$^4$ & 3.81x10$^4$ & 3.87x10$^4$ & 3.48x10$^5$ \\ \hline
        $k=2$ & 1.91x10$^4$ & 1.68x10$^4$ & 1.71x10$^4$ & 1.71x10$^4$ & 1.85x10$^4$ & 2.70x10$^4$  \\ \hline
        $k=4$ & 1.15x10$^4$ & 1.05x10$^4$ & 1.04x10$^4$ & 1.11x10$^4$ & 1.24x10$^4$ & 1.46x10$^4$  \\ \hline
        $k=8$ & 8.47x10$^3$ & 7.55x10$^3$ & 7.65x10$^3$ & 8.16x10$^3$ & 9.48x10$^3$ & 9.21x10$^3$ \\ \hline
        $k=16$ & 5.93x10$^3$ & 5.50x10$^3$ & 5.67x10$^3$ & 6.71x10$^3$ & 7.86x10$^3$ & 7.40x10$^3$ \\ \hline
        $k=32$ & 4.00x10$^3$ & 4.48x10$^3$ & 4.95x10$^3$ & 5.62x10$^3$ & 7.02x10$^3$ & 6.21x10$^3$ \\ \hline
    \end{tabular} \vspace{+10pt}
    \caption{Total compaction times for the \texttt{filluniquerandom} benchmark for different numbers of compaction threads ($k$) and fanouts ($F$) as well as our single level layout (S.L.). }
    \label{tab:insert}
    \vspace{-10pt}
\end{table}

We first consider query, scan, and compaction time with respect to fanout and concurrency.  
We use the benchmarks described above while varying $k$ from $1$ to $32$ and $F$ from $2$ to $32$, as well as our proposed single-level layout.
The results are shown in Tables \ref{tab:query} and \ref{tab:insert}.
RocksDB's standard layout results with varying $F$ and $k$ are plotted against our predictions using the MQSSD model assuming full concurrency in Figure \ref{fig:rocksdb}. 
Note that our predictions using the MQSSD model are not intended to be exact, but rather, they are meant for relative comparison and asymptotic analysis.  
The predictions in \ref{fig:rocksdb} use Equations \ref{eq:insert} and \ref{eq:query}, but we reintroduce constants to align our predictions with the experimental data.  

For all three workloads, we see in Figure \ref{fig:rocksdb} that the MQSSD model can predict the relative performance with respect to both $F$ and $k$; however, the accuracy of our compaction predictions diminish as $k$ increases. 
Compactions are generally more complex and variable, so we do not expect our predictions to be as accurate.
The decreased accuracy observed for larger $k$ is at least partly because we assume in our predictions that all threads can continuously perform concurrent I/O.  
Figure \ref{fig:rocksdb}A and \ref{fig:rocksdb}B suggest this assumption is reasonably accurate for queries and scans, provided sufficient user concurrency. 
By contrast, Figure \ref{fig:rocksdb}C suggests this is not true for RocksDB's pool of compaction threads. 
In particular, we see more minor deviations in performance than our prediction would suggest for $k \geq 8$. 
In addition, the observed compaction curves do not flatten with respect to $F$ as much as the model predicts as $k$ increases. 
We believe this due to the limitations in RocksDB's ability to support concurrent compaction as described in Section \ref{sec:insights}.
This is further supported by how the observed compaction performance is impacted by $F$. 
As discussed earlier, increasing $F$ increases write amplification per level, but if $F$ is too small, the number of levels begins to outweigh the reduction in write amplification per level.  
Looking at Figure \ref{fig:rocksdb}C, we can observe this trend for most $k$, but for large $k$, we see that $F=2$ starts to outperform most other values of $F$.
This is because the additional levels provide more opportunities for concurrent compactions.
However, the increased number of levels also significantly impacts read amplification, which scales with the number of levels.

We now compare the single-level layout to standard leveling to demonstrate how facilitating concurrency can expose new tradeoffs.  
As seen in Table \ref{tab:scan}, with standard leveling, RocksDB achieves its best query and scan performance across all values of $k$ when $F=32$.  
However, as we see in Table \ref{tab:insert} and Figure \ref{fig:rocksdb}C, $F=32$ also results in one of the worst write throughputs across all values of $k$. 
For most values of $k$, a fanout of $4$ or $8$ achieves the best write throughput.   
The standard LSM tree conventions suggest we use a layout with $F$ between 8 and 16, depending on the workload, to achieve a good balance between read and write amplification.  
However, if we consider how the layout affects concurrency, then there are new tradeoffs we can make.  
As shown in Table \ref{tab:insert}, for $k \geq 16$, the single-level layout achieves write throughput close to $F=16$ and better than $F=32$.  
At the same time, the single-level layout achieves zero read amplification and, therefore, outperforms RocksDB's standard leveling on queries and scans for all fanouts and any number of concurrent user threads, which can be seen in Table \ref{tab:query}. 
While this results in somewhat worse write throughput when compared to $F=16$, if RocksDB is configured using a large $F$, reads are already being prioritized. 
In such cases, the single-level layout's substantial improvement in read performance will likely be worthwhile. 
Another important consideration is that, even though the increased concurrency reduces the effect of write amplification on throughput, the increased write amplification will impact the SSD's lifespan. 
This configuration would then be ideal for low-write workloads or workloads that need to prioritize reads.  
For example, Laser is a high query throughput key-value storage service built on RocksDB that Facebook uses for real-time processing and requires low latency but only updates its data once per day \cite{realtime_rocks}.

\begin{figure}
  \begin{minipage}[b]{.55\linewidth}
    \centering
    \includegraphics[width=\linewidth]{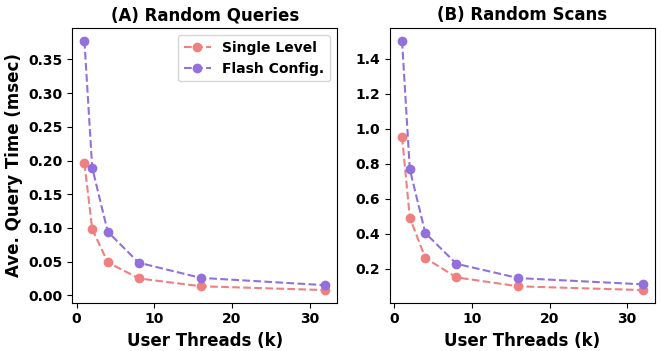}
    \captionof{figure}{RocksDB performance for random queries and scans comparing our layout optimized for multi-queue SSDs (Single Level) vs. a configuration used by Facebook for flash storage (Flash Config.) that uses $F=8$ and $C=8$. }
    \label{fig:flash_comp} 
  \end{minipage}\hfill \vspace{-5pt}
  \begin{minipage}[b]{.4\linewidth}
    \centering
    \begin{tabular}{|c||c|c||c|c|} \hline
        \multicolumn{5}{|c|}{Read Times for Flash Layouts (sec)} \\ \hline
          & \multicolumn{2}{c||}{Query} & \multicolumn{2}{c|}{Scan} \\ \hline 
          & S.L. & F.C. & S.L. & F.C. \\ \hline \hline
        $k=1$ & 627 & 1210 & 3050 & 4800 \\ \hline
        $k=2$ & 314 & 606 & 1570 & 2460 \\ \hline
        $k=4$ & 157 & 302 & 835 & 1290 \\ \hline
        $k=8$ & 80.6 & 155 & 481 & 730 \\ \hline
        $k=16$ & 43.2 & 83.0 & 317 & 468 \\ \hline
        $k=32$ & 25.2 & 48.4 & 249 & 354 \\ \hline

    \end{tabular} \vspace{+10pt}
    \captionof{table}{Total \texttt{db\_bench} scan and query times for our single-level layout (S.L.) vs. Facebook's deployed flash configuration (F.C.), which uses $F=8$, $C=8$, and $4$ compaction threads.  Total compaction times for the insert workload were 6210 \& 9330 sec. for the S.L. with $32$ compaction threads and F.C., respectively. }
    \label{tab:flash_comp}
  \end{minipage} \vspace{-10pt}
  
\end{figure}




Finally, we compare the single-level layout to an example configuration described in RocksDB's tuning guide that Facebook uses for flash storage \cite{rocksdb}. 
This configuration uses all the RocksDB defaults described earlier, except with $F=8$, $C=8$, and $4$ compaction threads.  
As described in the tuning guide, this configuration is used for workloads requiring primarily point queries and scans, so we use the same benchmarks. 
This description also suggests the single-level layout should be an ideal candidate despite its high write amplification. 
The results for both Facebook's flash configuration (F.C.) and our single-level layout (S.L.) are shown in Table \ref{tab:flash_comp} and Figure \ref{fig:flash_comp}. 
Despite being explicitly optimized for flash storage, the single-level layout achieves a better tradeoff by prioritizing concurrency and minimizing read amplification. 
This results in a roughly $2$x reduction in average point query time across all values of $k$. 
The improvement for scans is less pronounced since both configurations must still read the same total number of key-value pairs.  
Though write throughput is not the priority for this particular use case, a sufficient compaction thread pool can more than make up for the additional write amplification. 
Facebook's flash configuration required 9330 sec for compaction, while the single-level layout only required 6210 sec using 32 compaction threads.  
This illustrates how even systems optimized for flash storage may need to reconsider their performance tradeoffs when migrating to multi-queue SSDs. 

\section{Related Work}
\label{sec:related_work}

Other recent work has been related to modeling the performance of multi-queue SSDs, but it has primarily focused on emulation and simulation.  
Projects such as SimpleSSD \cite{simplessd}, Amber \cite{amber}, MQSim \cite{mqsim}, and MQSim-E \cite{mqsime} model the SSD in fine-grained detail at the architectural level.  
These simulators incorporate many individual models representing the state of the various components within the SSD. 
While these are immensely helpful for exploring the effects of architectural design choices, they are not ideal for software design, as running a full simulation to test a design can be quite computationally expensive and time-consuming.  
Furthermore, they provide a level of detail that can be opaque to those unfamiliar with SSD architectures. 

Other projects, like Cosmos+ \cite{cosmos} and FEMU \cite{femu}, have taken the route of emulation, which aims to fully replicate the operations of the SSD using either software or a configurable hardware platform.  
Even without the hardware platforms being prohibitively expensive, such emulators are intended as drop-in replacements for the SSD and require implementing and running software to gain insights.  

In contrast with both of these goals, we do not aim to achieve this level of fidelity in our analysis. 
Instead, we aim to provide a framework for informative comparisons at the earlier stages of software development that lends itself to formal reasoning (proofs) and lightweight approximations for online optimization.

\section{Conclusion \& Future Work}
\label{sec:conclusion}

This paper introduces the MQSSD model, a storage abstraction a storage abstraction tailored for algorithmic analysis of modern multi-queue SSDs.  
We review the hardware characteristics captured by the MQSSD model and highlight the limitations of previous I/O abstractions in addressing these features. 
We apply our model to Facebook's LSM-tree-based key-value store, RocksDB, and demonstrate that optimizing for modern storage hardware will require a shift in our thinking about I/O.  
Specifically, data layouts that enable higher levels of concurrent I/O are essential for fully leveraging the capabilities of modern storage hardware. Throughput gains from increased concurrency can also expose new performance tradeoffs that are better suited to particular applications.  
We conclude the MQSSD model provides a more accurate abstraction of modern hardware than previous models, allowing for greater insight and optimization when utilizing multi-queue SSDs.

In future work, we would like to extend this model to account for the interplay between reads and writes, as both operations ultimately utilize the same available bandwidth and impact each other's performance.  
Additional considerations include the impact of garbage collection as the volume of valid data grows, which can lead to increased write amplification.  
While these processes have been modeled in detail \cite{mqsim}, incorporating them without significantly increasing the model’s complexity remains challenging.  
While the MQSSD model is more complex than its predecessors, this added complexity is necessary to accurately represent modern hardware.

\bibliographystyle{ACM-Reference-Format}
\bibliography{works_cited}

\end{document}